\documentclass[sigconf]{acmart}
\usepackage{xcolor}
\usepackage{underscore}
\usepackage{enumitem}

\AtBeginDocument{%
  }

\copyrightyear{2025} 
\acmYear{2025} 
\setcopyright{acmlicensed}\acmConference[WWW Companion '25]{Companion Proceedings of the ACM Web Conference 2025}{April 28-May 2, 2025}{Sydney, NSW, Australia}
\acmBooktitle{Companion Proceedings of the ACM Web Conference 2025 (WWW Companion '25), April 28-May 2, 2025, Sydney, NSW, Australia}
\acmDOI{10.1145/3701716.3715865}
\acmISBN{979-8-4007-1331-6/2025/04}

\begin{document}

\title{Generative Large Recommendation Models: Emerging Trends in LLMs for Recommendation}

\author{Hao Wang}
\affiliation{
    \institution{University of Science and Technology of China}
  \city{Hefei}
  \state{Anhui}
  \country{China}
}
\authornote{Corresponding author.}
\email{wanghao3@ustc.edu.cn}

\author{Wei Guo}
\affiliation{
  \institution{Huawei Noah's Ark Lab}
  \state{Singapore}
  \country{Singapore}
}
\email{guowei67@huawei.com}

\author{Luankang Zhang}
\affiliation{
   \institution{University of Science and Technology of China}
  \city{Hefei}
  \state{Anhui}
  \country{China}
}
\email{zhanglk5@mail.ustc.edu.cn}

\author{Jin Yao Chin}
\affiliation{
  \institution{Huawei Noah's Ark Lab}
  \state{Singapore}
  \country{Singapore}
}
\email{chin.jin.yao@huawei.com}

\author{Yufei Ye}
\affiliation{
   \institution{University of Science and Technology of China}
  \city{Hefei}
  \state{Anhui}
  \country{China}
}
\email{aboluo2003@mail.ustc.edu.cn}

\author{Huifeng Guo}
\author{Yong Liu}
\affiliation{
  \institution{Huawei Noah's Ark Lab}
  \country{Shenzhen, China \& Singapore}
}
\email{huifeng.guo@huawei.com}
\email{liu.yong6@huawei.com}

\author{Defu Lian}
\affiliation{
   \institution{University of Science and Technology of China}
  \city{Hefei}
  \state{Anhui}
  \country{China}
}
\email{liandefu@ustc.edu.cn}

\author{Ruiming Tang}
\affiliation{
  \institution{Huawei Noah's Ark Lab}
  \state{Shenzhen}
  \country{China}
}
\email{tangruiming@huawei.com}

\author{Enhong Chen}
\affiliation{
   \institution{University of Science and Technology of China}
  \city{Hefei}
  \state{Anhui}
  \country{China}
}
\email{cheneh@ustc.edu.cn}

\renewcommand{\shortauthors}{Wang et al.}

\begin{abstract}
In the era of information overload, recommendation systems play a pivotal role in filtering data and delivering personalized content. Recent advancements in feature interaction and user behavior modeling have significantly enhanced the recall and ranking processes of these systems. With the rise of large language models (LLMs), new opportunities have emerged to further improve recommendation systems. This tutorial explores two primary approaches for integrating LLMs: LLMs-enhanced recommendations, which leverage the reasoning capabilities of general LLMs, and generative large recommendation models, which focus on scaling and sophistication. While the former has been extensively covered in existing literature, the latter remains underexplored. This tutorial aims to fill this gap by providing a comprehensive overview of generative large recommendation models, including their recent advancements, challenges, and potential research directions. Key topics include data quality, scaling laws, user behavior mining, and efficiency in training and inference. By engaging with this tutorial, participants will gain insights into the latest developments and future opportunities in the field, aiding both academic research and practical applications. The timely nature of this exploration supports the rapid evolution of recommendation systems, offering valuable guidance for researchers and practitioners alike.
\end{abstract}



\maketitle

\section{FORMAT AND INTENDED AUDIENCE}
\begin{itemize}[left=0pt]
    \item \textbf{Format.} This tutorial will be a 3-hour, on-site session conducted in a lecture-style format. Presenters will attend in person to facilitate direct interaction and engagement with participants. The session will include a mix of presentations and interactive discussions, allowing for questions and feedback throughout.

    \item \textbf{Intended Audience.} This tutorial is designed for researchers and academics interested in the latest advancements in generative recommendation models. Participants should have a basic understanding of large language models and traditional recommendation systems. The session will explore opportunities, developments, and challenges in generative recommendation models, providing insights into future trends. It is particularly beneficial for information retrieval professionals aiming to deepen their expertise and explore new research directions. Additionally, professionals from other fields can gain strategic insights for designing domain-specific generative models.
\end{itemize}

\section{PRESENTERS}

\textbf{Hao Wang}\footnote{\url{http://staff.ustc.edu.cn/~wanghao3/}} is an Associate Researcher at the University of Science and Technology of China (USTC), specializing in representation learning, graph mining, sequential modeling, and LLMs.

\textbf{Wei Guo}\footnote{\url{https://scholar.google.com/citations?user=9NGhGNgAAAAJ&hl=zh-CN}} is a Researcher at Huawei Noah's Ark Lab with an M.Sc. from Wuhan University. His research interests include recommender systems, deep learning, and data mining.

\textbf{Luankang Zhang}\footnote{\url{https://scholar.google.com/citations?user=O5Ib6NYAAAAJ&hl=zh-CN&oi=ao}} is a Ph.D. student at USTC, focusing on recommender systems and data mining.

\textbf{Jin Yao Chin}\footnote{\url{https://scholar.google.com/citations?user=_CxUOZ4AAAAJ&hl=en}} is a Researcher at Huawei Noah's Ark Lab. His research focuses on recommender systems and deep learning.

\textbf{Yufei Ye}\footnote{\url{https://orcid.org/0009-0009-1193-0426}} is an undergraduate at USTC, researching recommendation systems under Dr. Hao Wang and Prof. Defu Lian.

\textbf{Huifeng Guo}\footnote{\url{https://scholar.google.com/citations?user=jlBcPn8AAAAJ&hl=zh-CN}} is a Senior Researcher at Huawei Noah's Ark Lab. He specializes in recommendation, information retrieval, and advertising, focusing on deep learning models.

\textbf{Yong Liu}\footnote{\url{https://scholar.google.com/citations?user=IY24geQAAAAJ&hl=zh-CN&oi=sra}} is a Senior Principal Researcher at Huawei Noah's Ark Lab. His research includes large language models, search, and recommendation systems.

\textbf{Defu Lian}\footnote{\url{https://faculty.ustc.edu.cn/liandefu/zh_CN/index/988651/list/index.htm}} is a Professor at USTC, focusing on data mining, deep learning, and recommender systems. He has published extensively and received awards for his research.

\textbf{Ruiming Tang}\footnote{\url{https://scholar.google.com.sg/citations?user=fUtHww0AAAAJ&hl=en}} is the Director of the Recommendation \& Search Lab at Huawei Noah's Ark Lab. His research areas include recommender systems and deep learning. He has published over 100 papers and reviews for top conferences like TKDE and KDD.

\textbf{Enhong Chen}\footnote{\url{https://cs.ustc.edu.cn/2020/0806/c23235a460077/pagem.htm}} is a Professor and executive dean at USTC. He is a CCF Fellow and IEEE Fellow.

\section{TOPIC AND RELEVANCE}
\subsection{Motivation}
In today's era of information overload, recommendation systems~\cite{xie2024breaking,xie2024bridging,shen2024exploring,yin2024learning,han2024efficient,wang2025mf,yin2023apgl4sr,han2023guesr,zhang2022clustering,wang2021hypersorec,wang2019mcne,zhang2024unified} have become essential tools for filtering vast amounts of data and delivering personalized content to users based on their historical interactions. Over the past decade, significant advancements have been made in the field, particularly in techniques such as feature interaction~\cite{wang2021dcn,guo2017deepfm} and user behavior modeling~\cite{zhou2018deep,zhou2019deep,xu2024multi}. These innovations have led to substantial improvements in the key processes of recommendation systems, namely recall and ranking, enhancing their ability to accurately predict and present information that aligns with users' interests.

Recently, large language models (LLMs)have achieved remarkable success and show great potential for application in recommendation systems. There are currently two primary approaches to integrating LLMs into these systems. The first approach involves enhancing traditional recommendation systems by utilizing the extensive world knowledge and reasoning capabilities of general LLMs, a method known as \textbf{LLMs-enhanced recommendation}~\cite{xi2024towards,liu2024once,li2023ctrl}. The second approach focuses on developing larger and more sophisticated recommendation models by designing generative recommendation models that adhere to scaling laws, known as \textbf{generative large recommendation models}~\cite{zhang2024wukong,zhang2024scaling,zhai2024actions}. In recent months, there have been groundbreaking advancements in this second path, highlighted by the successful online deployment of trillion-parameter generative recommendation models. These models, comparable in scale to GPT-3 and LLaMa-2, have achieved significant business improvements and demonstrated powerful scalability, attracting considerable attention from both academia and industry. Despite advancements, previous surveys~\cite{lin2023can,wu2024survey} and tutorials~\cite{zhang2024large,liu2023user} on LLMs for recommendation have mainly focused on the first approach, leaving a gap in the systematic exploration of the second path. Therefore, this tutorial aims to offer a comprehensive summary of the first approach and, for the first time, systematically introduce the advancements of the second approach. This dual focus makes the tutorial both necessary and timely.

In addition to an in-depth analysis of the paradigm shift in modeling and the evolution of technical directions within the field of LLMs for  recommendation, this tutorial will also address the key challenges and issues generative large recommendation models face, alongside presenting our preliminary explorations. The tutorial will begin by focusing on the critical issue of data quality, highlighting data-centric research and data regeneration methods for large models. Next, it will examine how data size and model size influence recommendation performance, exploring the scaling laws and performance metrics pertinent to the recommendation domain. Additionally, the tutorial will delve into methods for more effectively mining and utilizing complex user behavior sequences to enhance data usage. Finally, it will address the challenge of improving training and inference efficiency, which is crucial for enabling the scalability of model size.

This tutorial will conclude by exploring future research directions in the field of generative large recommendation models. One promising direction is conducting data engineering tasks, such as data distillation and data selection, to ensure that training data is clean, relevant, and well-structured. At the representation level, employing tokenizer technology and integrating side information can significantly enhance recommendation performance and efficiency. Additionally, addressing incremental updates in practical applications presents a valuable research opportunity. By engaging with this tutorial, participants will gain insights into the latest advancements in LLMs for recommendations, develop an understanding of the current challenges in the field, and identify future research directions. This knowledge will aid participants in applying or further researching generative large recommendation models.

\textbf{Necessity and timely of this tutorial.}  Large models for recommendation have gained significant attention recently, with many publications and tutorials at conferences like WWW'24 \cite{zhang2024large}. However, they focus on LLMs-enhanced recommendations, leaving a gap in exploring generative large recommendation models. With their recent successes, a comprehensive tutorial is crucial to support the field's rapid development, offering valuable insights for practitioners. Therefore, this tutorial is both necessary and timely.

\subsection{Objective}

The objective of this tutorial is to provide an in-depth exploration of large language models (LLMs) for recommendation systems, with a particular focus on addressing the gap in the systematic study of generative large recommendation models. This tutorial will cover the latest developments, existing challenges, and potential future directions in this area. Additionally, it aims to equip participants with the skills and knowledge necessary to harness generative large recommendation models in their projects and to contribute to advancements in related fields.

\subsection{Relevance}

This tutorial is highly relevant to the Web Conference themes of user modeling, personalization, and recommendation. Previous tutorials, such as those at RecSys'23 \cite{hua2023tutorial} and WWW'24 \cite{zhang2024large}, have explored using large language models (LLMs) to enhance traditional recommendation systems, focusing mainly on their applications. In contrast, our tutorial not only summarizes the latest advancements in LLM-enhanced recommendations but also systematically introduces, for the first time, the design of generative large recommendation models that adhere to scaling laws. Our content is broader, more cutting-edge, and addresses a wider range of challenges than previous tutorials. Given its strong relevance to the Web Conference, the significant recent achievements in this field, and the comprehensive and innovative nature of our tutorial, we anticipate it will attract considerable attention.

\subsection{Outline}
The following outline provides a detailed overview of the topics and structure of this tutorial:

\begin{itemize}[left=0pt]
    \item \textbf{Introduction (\textcolor{red}{15min, Enhong Chen, Ruiming Tang})}
    \begin{itemize}
        \item Overview of the tutorial structure and objectives
        \begin{itemize}
            \item Introduction of speakers and their expertise
        \end{itemize}
        \item Background on recommendation systems (RS)
        \item Technological advantages of large language models (LLMs)
        \item Potential of integrating LLMs with RS
    \end{itemize}
    
    \item \textbf{Large Language Models for Recommendation (\textcolor{red}{45min, Huifeng Guo, Yong Liu})}
    \begin{itemize}
        \item LLMs-enhanced recommendation
        \begin{itemize}
            \item LLMs embeddings + recommendation~\cite{li2023ctrl}
            \item LLMs tokens + recommendation~\cite{liu2024once,xi2024towards}
            \item LLMs as recommendation systems~\cite{zhai2023knowledge}
        \end{itemize}
        \item Generative large recommendation models~\cite{zhang2024scaling,zhai2024actions,liu2024multi,chen2024hllm,ye2025fuxi}
    \end{itemize}
    
    \item \textbf{\textcolor{red}{Q\&A Session (5 min)}}
    
    \item \textbf{\textcolor{red}{Break (10 min)}}
    
    \item \textbf{Issues, Challenges, and Preliminary Exploration of Generative Large Recommendation Models (\textcolor{red}{60min, Luankang Zhang, Jin Yao Chin, Yufei Ye})}
    \begin{itemize}
        \item Data-centric paradigm~\cite{yin2024dataset,zhang2025td3,yin2024entropy}
        \item Scaling laws
        \begin{itemize}
            \item Changes in loss with data and model scaling~\cite{niu2024beyond}
            \item Performance laws~\cite{du2024understanding, niu2024beyond}
        \end{itemize}
        \item Complex user behavior modeling
        \begin{itemize}
            \item Long sequence modeling~\cite{ren2019lifelong, si2024twin, yu2024ifa}
            \item Multi-behavior modeling~\cite{wu2022multi, liu2024multi}
            \item Multi-domain behavior modeling~\cite{park2024pacer, liu2024graph}
        \end{itemize}
        \item Efficiency optimization~\cite{xu2023efficient, shao2024one, zhai2023bytetransformer, holmes2024deepspeed}
    \end{itemize}
    
    \item \textbf{Conclusions and Future Directions (\textcolor{red}{40min, Hao Wang, Wei Guo})}
    \begin{itemize}
        \item Summary of key points
        \item Future research directions
        \begin{itemize}
            \item Data engineering~\cite{lai2024survey, zhang2023sled, wang2023gradient}
            \item Representation enhancement
            \begin{itemize}
                \item Tokenizer application~\cite{hou2023learning, wang2024learnable}
                \item Side information modeling~\cite{he2023survey, liu2023user}
            \end{itemize}
            \item Incremental learning~\cite{ song2024increasing}
        \end{itemize}
    \end{itemize}
    
    \item \textbf{\textcolor{red}{Q\&A Session (5 min)}}
\end{itemize}

\subsection{Qualification of presenters}
Our team has extensive experience in the field of recommender systems, demonstrated by numerous publications in prestigious conferences such as WWW~\cite{lu2023differentiable, han2024efficient, wan2022cross}. 
We have also organized tutorials and workshops, such as the tutorial at RecSys'23~\cite{liu2023user} and the workshops at RecSys'23\footnote{\url{https://dlp4rec.github.io/}} and CIKM'23\footnote{\url{https://rgm-cikm23.github.io/}}. Notably, our recent work~\cite{yin2024dataset} was recognized with the KDD Best Student Paper Award for its innovative data-centric approach to enhancing sequential recommender systems. Additionally, our survey~\cite{wu2024survey} is the first comprehensive review of the integration of large language models (LLMs) in recommendation systems. Our technical report~\cite{guo2024scaling} offers pioneering insights into generative large recommendation models. We also explore the theoretical aspects of large recommendation models, examining the scaling laws between data compression rate, quality, and model performance~\cite{shen2024predictive}.
Our profound expertise and pioneering achievements equip us to deliver a thorough and insightful presentation on these cutting-edge developments~\cite{zhang2024learning,zhang2020context,zhang2019graph,zhang2022hierarchical,zhang2022cglb,yang2024exploring,li2024configure,yang2023lever,penglive} .

\section{TUTORIAL DETAILS}

\begin{itemize}[left=0pt]
    \item \textbf{Duration.} This tutorial is designed to span a period of 3 hours, allowing for comprehensive coverage of the topic.
    
    \item \textbf{Tutorial Materials.} Slides will be posted on the tutorial website, and organizers can obtain necessary copyright permissions.
    
    \item \textbf{Video Teaser.} A 3-minute video teaser summarizing the tutorial content and objectives is available on Google Drive \footnote{\url{https://drive.google.com/file/d/1NSgFMsthh18msf1wdU7Tt1d7iXtK8pDG/view?usp=sharing}}. 
    
    \item \textbf{Organization Details.} The tutorial will be conducted on-site, with all presenters attending in person. In the event of unforeseen circumstances, we are prepared to offer pre-recorded lectures. We are also open to live streaming the tutorial on popular platforms, pending approval, to enhance remote accessibility.
\end{itemize}

\bibliographystyle{ACM-Reference-Format}
\bibliography{Sections/References}

\end{document}